\documentclass[aps,twocolumn]{revtex4}
\usepackage{epsfig}
\usepackage{graphicx}
\newcommand\ket[1]{\left|#1\right\rangle}
\newcommand\bra[1]{\left\langle#1\right|}
\newcommand\braket[2]{\left\langle#1\right|\left.\!\!#2\right\rangle}

\newcommand\so{{\hat\sigma}}
\newcommand\mo{{\hat\mu}}
\newcommand\re{{\rm Re}}

\begin{document}

\title{Test of weak measurement on a two- or three-qubit
computer}
\author{Todd A. Brun}
\affiliation{Communication Sciences Institute, University of Southern California, \\
Los Angeles, California 90089-2565, USA}
\author{Lajos Di\'osi}
\affiliation{Centro Internacional de Ciencias, 62131 Cuernavaca, Mexico}
\affiliation{Research Institute for Particle and Nuclear Physics\\
    H-1525 Budapest 114, P.O.Box 49, Hungary\footnote{permanent address}}
\author{Walter T. Strunz}
\affiliation{Centro Internacional de Ciencias, 62131 Cuernavaca, Mexico}
\affiliation{Institut f\"ur Theoretische Physik, Technische Universit\"at, 
Dresden, D-01062 Dresden, Germany\footnote{permanent address}}

\date{\today}
\begin{abstract}
Current quantum computer technology is sufficient to realize 
weak measurements and the corresponding concept of weak values.
We demonstrate how the weak value anomaly can be tested, along
with consistency and simultaneity of weak values, using only
discrete degrees of freedom. All you need is
a quantum computer with two---or better, three---qubits.
We also give an interpretation of the weak value as an
effective field strength in a postselected spin measurement.
\end{abstract}
\pacs{03.65.Ta 03.67.-a}
\maketitle

\section{Introduction}

Of the many seeming paradoxes of quantum mechanics, one of the most 
interesting and bizarre is the idea of a {\it weak value}.  First proposed 
by Aharonov, Albert and Vaidman \cite{AAV}, this uses a combination of 
weak measurements and postselection to derive measurement ``results''
which are far outside the normal range of values for the measured observable.
While a remarkable theoretical result, direct experiments that actually
demonstrate it are difficult---see, however the interpretation of correlation
functions as ``weak values'' by Wiseman \cite{wiseman2002} and the
quantum optical experiments described in \cite{foster2000} and \cite{pryde2005}.
With the rapid experimental progress from the surge of interest in 
quantum information processing, it may be possible to do such experiments 
in a highly controlled, repeatable fashion, using only discrete degrees of 
freedom.

In quantum measurements, the act of acquiring information about a quantum 
system is always accompanied by a complementary disturbance of the system.  
This is the content of the famous uncertainty principle of Heisenberg.  
A measurement which does not change the state of the system must also yield 
no information.

It is possible in principle, however, to make the disturbance as small as 
one likes, so long as one is content to acquire correspondingly little 
information.  This is the idea of a weak measurement.  To perform such a 
measurement in practice, one must generally cause the system to interact 
weakly with a second system---an ancillary system, or {\it ancilla}, 
sometimes called the ``meter''---which is under one's experimental control, 
and has been prepared in a known initial state.  This ancilla then 
undergoes a strong measurement of its own.  In the limit where the system 
and ancilla do not interact at all, clearly this measurement will yield 
no information.  As we gradually increase the strength of the interaction, 
the measurement outcome will contain more and more information about the 
system, until eventually the effect is the same as performing a strong 
measurement directly on the system.

The idea of postselection supposes that instead of performing repeated 
measurements on a single system, one prepares many copies of the 
system by repeating the same preparation procedure over and over.  These 
copies will all have the same initial state.  These copies then undergo 
some standard operation---some sequence of unitary transformations and 
measurements---followed by a final measurement.  One then keeps the data 
only from those systems whose final measurement gave a particular outcome, 
and averages results over this sub-ensemble.

In this paper we will review the Aharonov, Albert, Vaidman definition of 
weak values, and then describe how experimental systems designed for 
quantum computation can lead to an immediate experimental implementation
using existing quantum computers, for example in ion trap quantum computers
\cite{schmidtkaler2003,seidelin06,zhu2006}.

\section{Two Qubit indirect measurement device}

A qubit is a two-dimensional quantum system, with a standard 
(``computational'') basis which we denote $\{\ket0,\ket1\}$.  There can be 
many different physical embodiments of such a system:  the spin of an 
electron, the polarization of a single photon, a two-level subspace of the 
electronic states of an atom or ion, etc.  For quantum algorithms, much work 
has been devoted to the performance of {\it quantum gates}, analogous to 
classical logic gates, which effect a unitary transformation of one or two 
qubits at a time.  The canonical two-qubit gate is the {\it controlled-NOT} 
(CNOT):
\begin{equation}\label{CNOT}
\ket{i}\otimes\ket{j} \longrightarrow
U_{SA} \left[\ket{i}\otimes\ket{j}\right]
= \ket{i}\otimes\ket{j\oplus i}~,
\end{equation}
where $j\oplus i$ is the exclusive-OR (XOR) of the bit values $i$ and $j$,
and $U_{SA}$ is the unitary transformation which represents a CNOT
between the system and the ancilla.

A quantum circuit with a single CNOT gate makes a perfectly controllable 
indirect measurement of a qubit in the computational basis, by storing 
the value of the qubit in a second (ancilla) qubit:
\[
\ket{i}_S \otimes\ket{0}_A \longrightarrow \ket{i}_S\otimes\ket{i}_A ,\qquad
(i=0,1),
\]
where $S$ and $A$ label the system and ancilla, respectively, the system 
is initially in the computational state $\ket{i}$, and the ancilla is 
initially in the state $\ket{0}$.  (We will suppress the labels $S,A$ where 
there is no possibility of confusion.)  The ancilla can then be measured 
by a strong measuring device, which will simultaneously ``collapse the 
wavefunction'' of the system qubit.  This type of indirect measurement 
can be very useful when the only direct measurements are destructive 
(for example, a photodetector which absorbs the photon it is measuring).  
If the system is initially in a superposition of computational basis states, 
it will become entangled with the ancilla:
\[
(\alpha\ket0+\beta\ket1) \otimes\ket{0} \longrightarrow 
\alpha\ket0\otimes\ket0 + \beta\ket1\otimes\ket1 .
\]
When the ancilla is measured, one of these two terms will be selected with 
probability $|\alpha|^2$ or $|\beta|^2$.

Suppose now that instead of $\ket0$ we prepare the ancilla in the initial
superposition 
\[
\ket{\psi_A} = \cos\frac{\vartheta}{2}\ket{0}+\sin\frac{\vartheta}{2}\ket{1}.
\]
Let the system qubit be in the state 
$\ket{\phi_i}=\alpha\ket{0} + \beta\ket{1}$, and have the two qubits interact 
via the CNOT. Then we measure the ancilla in its computational basis in order 
to obtain information about $\ket{\phi_i}$. After the CNOT, the system and 
ancilla are in the state
\begin{equation}
\ket{\phi_i}\otimes\ket{\psi_A} \longrightarrow
  \ket{\Psi} = U_{SA}[\ket{\phi_i}\otimes\ket{\psi_A}]
\end{equation}
with
\begin{eqnarray}
\ket{\Psi} & = &
\alpha \ket0 \otimes(\cos\frac{\vartheta}{2}\ket{0}+
                       \sin\frac{\vartheta}{2}\ket{1}) \nonumber\\
&& + \beta \ket1 \otimes(\sin\frac{\vartheta}{2}\ket{0}+
                         \cos\frac{\vartheta}{2}\ket{1}) .
\label{post_interaction}
\end{eqnarray}
If $\vartheta=0$ then $\ket{\psi_A}=\ket0$ and this is the case we have just 
considered:  the indirect measurement is perfectly equivalent with a direct 
measurement of the first qubit. If $\vartheta=\pi/2$ then the indirect 
measurement does not give any information on the first qubit, whose state 
$\ket{\phi_i}$ will just survive the procedure unchanged, without being 
entangled with the state of the ancilla. Hence, the parameter $\vartheta$ 
offers full control of the strength of the indirect measurement. 
We shall be interested in weak measurements, which are realized by 
$\vartheta=(\pi/2)-\epsilon$ where $0<\epsilon\ll 1$. This will be 
discussed later.

Let us determine the expectation value of the operator 
$\so_z \equiv \ket0\bra0 - \ket1\bra1$ of the ancilla in the state $\ket{\Psi}$
given by Eq.~(\ref{post_interaction}):
\begin{eqnarray}
\langle\so_z^{\rm{ancilla}}\rangle &=& |\alpha|^2 \cos^2\frac{\vartheta}{2}
  + |\beta|^2\sin^2 \frac{\vartheta}{2} \nonumber\\
&& - |\alpha|^2 \sin^2\frac{\vartheta}{2}
  - |\beta|^2\cos^2 \frac{\vartheta}{2} \nonumber\\
&=& (|\alpha|^2-|\beta|^2) \left(\cos^2\frac{\vartheta}{2} - 
                             \sin^2\frac{\vartheta}{2} \right) \nonumber\\
&=& (|\alpha|^2 - |\beta|^2) \cos\vartheta .
\end{eqnarray}
Since $|\alpha|^2-|\beta|^2=\bra{\phi_i}\so_z\ket{\phi_i}$ (which we
simply denote by $\langle\so_z\rangle$), it follows that
\begin{equation}
\langle\so_z\rangle = \frac{1}{\cos\vartheta}\langle\so_z^{\rm{ancilla}}\rangle
\label{ancilla_expectation}
\end{equation}
where the expectation value on the l.h.s. is the expectation value
of $\so_z$ in the system initial state $\ket{\phi_i}$, while the expectation 
value on the r.h.s. stands for the post-interaction expectation value of 
$\so_z^{\rm{ancilla}}$.

The simple relationship (\ref{ancilla_expectation}) suggests that we can still measure the system
expectation value of $\so_z$ if we measure the ancilla expectation value
of $\so_z^{\rm{ancilla}}$ instead, and rescale the result by $1/\cos\vartheta$.  Of course,
the statistical error of the indirect measurement is larger then the statistical
error of the direct measurement.  Suppose that many copies of the system
and ancilla are prepared in the same initial state.  For each copy, the CNOT
interaction is performed, and then the ancilla is measured in the
computational basis.  These measurements are used to estimate the
expectation value of the operator $\so_z^{\rm ancilla}$.  This latter quantity
is what we estimate from the measurement statistics:
\begin{equation}
\langle\so_z^{\rm{ancilla}}\rangle\approx\frac{N_{0}-N_{1}}{N_{0}+N_{1}}
\label{s_ancilla_N}
\end{equation}
where $N_0, N_1$ are the measurement counts corresponding to
the outcomes $\ket0$ and $\ket1$ when measuring $\so_z^{\rm{ancilla}}$, 
respectively, obtained from a total number of measurements $N=N_{0}+N_{1}$. 
Let us determine the statistical error of the quantity (\ref{s_ancilla_N}) for large $N$:
\begin{equation}
\Delta\langle\so_z^{\rm{ancilla}}\rangle\approx
 \sqrt{\frac{2(1+\langle\so_z^{\rm ancilla}\rangle)}{N}} ,
\label{Ds_ancilla_N}
\end{equation}
yielding the following statistical error of the indirect measurement of 
$\langle\so_z\rangle$:
\begin{equation}
\Delta\langle\so_z\rangle\approx\frac{1}{\cos\vartheta}
 \sqrt{\frac{2(1+\cos\vartheta\langle\so_z\rangle)}{N}} ,
\label{Ds_N}
\end{equation}
which increases with $\vartheta$. Observe that the value $\vartheta=0$ would formally 
correspond to the direct measurement. 

We are interested in the weak measurement limit $\vartheta=(\pi/2)-\epsilon$. 
To leading order in the small parameter $\epsilon$ we have:
\begin{equation}
\langle\so_z\rangle=
      \frac{1}{\epsilon}\langle\so_z^{\rm{ancilla}}\rangle,
\label{weak_value_eps}
\end{equation}
and
\begin{equation}
\Delta\langle\so_z\rangle\approx \frac{1}{\epsilon}\sqrt{\frac{2}{N}},
\label{Dweak_value_eps}
\end{equation}
since $\epsilon\ll1$. The latter equation also means that the statistical error
of a {\it single} measurement is $\sim\sqrt{2}/\epsilon$.
The two equations (\ref{weak_value_eps},\ref{Dweak_value_eps}) assure,
that our indirect measurement is a weak measurement of the system's $\so_z$, 
cf. the general definitions in \cite{Dio}:
i) our measurement yields the unbiased mean of $\langle\so_z\rangle$ 
and ii) the statistical error (regarding $\langle\so_z\rangle$) of a single measurement 
is much larger than total range of all possible values of the measured quantity 
$\so_z$. For an arbitrary small $\epsilon$ we need to have suitably large 
statistics $N\sim\epsilon^{-2}$ to yield an estimate of 
$\langle \so_z\rangle$ with any desired precision.
Accordingly, our weak measurement reproduces all basic features of the AAV weak measurement,
that we are going to show by detailed proofs in the forthcoming sections. 

Other than convenience, there is no particular reason to chose $\so_z$
as the observable. For later reference we mention that such an indirect 
measurement of, for instance, $\so_x$ is best formulated in terms
of its eigenstates $\ket\pm = (\ket0 \pm \ket1)/\sqrt{2}$. Accordingly,
the ancilla qubit should be prepared in the state
\begin{equation}
\ket{\psi^x_A} = \cos\frac{\vartheta}{2}\ket{+}+\sin\frac{\vartheta}{2}\ket{-}
\label{initial_ancilla_x}
\end{equation}
and for the corresponding CNOT$^x$ operation one has to replace the 
computational basis states $\{\ket0,\ket1\}$ by $\{\ket+,\ket-\}$ in expression
(\ref{CNOT}).  (This interaction is the same as the usual CNOT with the
control and target qubits interchanged.)

\section{Two qubit indirect measurement with postselection
and the weak value anomaly}

Up to this point, we have assumed that after the interaction with the 
ancilla and the ancilla's subsequent measurement we make no further use of 
the original system.  It is possible, however, to measure the system as well 
as the ancilla. Then, instead of the usual statistics including all measurement 
outcomes as described above, we keep only results where the additional system 
measurement confirms the system qubit to be in a certain final state
$\ket{\phi_f}$.  This is the idea of  {\it postselection}, described by Aharonov,
Albert and Vaidman in \cite{AAV}.  

Naively, we calculate the same quantity as before,
\begin{equation}
\frac{N_{f0}-N_{f1}}{N_{f0}+N_{f1}},
\label{postselection_rate}
\end{equation}
and we call it the {\it postselected estimate} of $\so_z^{\rm{ancilla}}$,
with respect to the final system state $\ket{\phi_f}$. As before, we
rescale the above quantity by $1/\cos\vartheta$ and expect that in the
large $N$ limit we obtain something sensible in terms of the system's
$\so_z$ and of the initial as well of the final states 
$\ket{\phi_i},\bra{\phi_f}$.
While this expectation fails in general, it becomes true in the weak measurement limit. 
Then, surprisingly, the postselection rate is just $\vert\bra{\phi_f}\phi_i\rangle\vert^2$,
independent of the (weak) interaction with the ancilla.
As  first defined by AAV \cite{AAV}, the so-called {\it weak value} 
of $\so_z$ is
\begin{equation}
{}_f\langle\so_z\rangle_i\equiv
 {\rm Re}\frac{\bra{\phi_f}\so_z\ket{\phi_i}}{\bra{\phi_f}\phi_i\rangle},
\end{equation}
(once again expressed in terms of the system qubit state). We will now show that in the large $N$ limit
\begin{equation}
\frac{1}{\epsilon}\frac{N_{f0}-N_{f1}}{N_{f0}+N_{f1}}\approx 
          {}_f\langle\so_z\rangle_i~.
\label{postselected_estimate}
\end{equation}
In other words, our indirect device {\it with} postselection measures the weak value of the system qubit, in the very same way that it gave us the ordinary value $\langle\so_z\rangle$ {\it without} postselection, c.f., Eq.~(\ref{weak_value_eps}).

Let us begin with the post-interaction state $\ket{\Psi}$ from 
(\ref{post_interaction}). The probabilities of finding the system in the 
state $\ket{\phi_f}$ and the ancilla in the state $\ket0$ or $\ket1$, 
respectively, are
\begin{eqnarray}
p_{f0} &=& \left| \bra{\phi_f}\left(\alpha\cos\frac{\vartheta}{2}\ket0 
        + \beta\sin\frac{\vartheta}{2}\ket1 \right) \right|^2 ,\nonumber\\
p_{f1} &=& \left| \bra{\phi_f}\left(\alpha\sin\frac{\vartheta}{2}\ket0 
        + \beta\cos\frac{\vartheta}{2}\ket1 \right) \right|^2  .
\label{joint_probs}
\end{eqnarray}
In the limit of large $N$, the postselected estimate of $\so_z^{\rm ancilla}$ 
conditioned on the outcome of the system measurement being $\ket{\phi_f}$ is 
$(p_{f0}-p_{f1})/(p_{f0}+p_{f1})$.  Unlike the case we considered in Sec.~II, 
in general these quantities have nothing universal to do with $\so_z$ of 
the system qubit.

However, the case becomes positive in the weak measurement limit 
$\vartheta = \pi/2 - \epsilon$.  In this limit there is indeed a relationship 
between the outcomes of the postselected measurement and $\so_z$ of the 
system qubit, which we can see by expanding Eqs.~(\ref{joint_probs}) to 
first order in $\epsilon$:
\begin{eqnarray}
p_{f0} &\approx& \frac{1}{2} \left( |\bra{\phi_f}\phi_i\rangle|^2 
    + \epsilon\cdot \re[\bra{\phi_f}\so_z\ket{\phi_i}\bra{\phi_i}\phi_f\rangle]
          \right),\nonumber\\
p_{f1} &\approx& \frac{1}{2} \left( |\bra{\phi_f}\phi_i\rangle|^2 
    - \epsilon\cdot\re[\bra{\phi_f}\so_z\ket{\phi_i}\bra{\phi_i}\phi_f\rangle] 
          \right)  .
\label{weak_probs}
\end{eqnarray}
To lowest order in $\epsilon$ we thus find
\begin{equation}
\frac{p_{f0} - p_{f1}}{p_{f0} + p_{f1}} 
 \approx \epsilon\cdot {\rm Re}\frac{\bra{\phi_f}\so_z\ket{\phi_i}}
{\bra{\phi_f}\phi_i\rangle} 
= \epsilon\cdot {}_f\langle\so_z\rangle_i.
\label{postselected_probabilities}
\end{equation}
If we repeat this procedure $N$ times and 
get $N_{f0}$ results $\ket{\phi_f}\ket0$ and $N_{f1}$ results 
$\ket{\phi_f}\ket1$, where $N_{f0}/N_{f1}\approx p_{f0}/p_{f1}$ holds, the
postselected indirect estimate of $\so_z$ (i.e. the weak value of $\so_z$) is just given by 
(\ref{postselected_estimate}) in the limit of large $N$, as claimed above.

Let us emphasize that we are extending the original AAV theory, which
makes use of a von~Neumann measuring device where the ancilla is a
fictitious particle, whose position serves as the meter.  In the case presented
here, the ancilla is just a qubit, and all of the degrees of freedom are
discrete. Such an extended theory of weak measurement was given
recently in \cite{Dio}, and a simple version of this was earlier used in \cite{Brun2002}.

The weak value anomaly is reflected in the fact that
if we invariably trust in our weak measurement device on a post-selected
ensemble as well as on the whole ensemble then the
postselected indirect estimate of $\so_z$ falls well outside the range $[-1,1]$ of 
``normal'' expectation values for $\so_z$.  For a concrete example,
consider the initial and final states
\begin{eqnarray}
\ket{\phi_i} & = & \frac{1}{\sqrt{2}}\left(\ket0 + \ket1\right) \nonumber \\
\ket{\phi_f} & = & \frac{1}{\sqrt{2(z^2+1)}}\left((z+1)\ket0-(z-1)\ket1\right)
\label{i_f_example}
\end{eqnarray}
with arbitrary real parameter $z$. 
We find
${}_f\langle\so_z\rangle_i=z$, and therefore the 
postselected estimate of $\langle\so_z\rangle$
would be $z$, which can take any (arbitrarily large) value.
The effect of the anomalous large mean value in the post-selected states is real: 
a probe will sense it as a large mean field, cf. Sec.~V.
Clearly, the more ``anomalous'' the
postselected estimate, i.e. the larger $z$, the less likely the
postselection criterion will be met:  the initial and final states
$\ket{\phi_i}$ and $\ket{\phi_f}$ are almost orthogonal, and the
probability for a successful run of the experiment is
$p_f = p_{f0} + p_{f1} \approx 1/(z^2+1)$.  For large $z$,
most runs of the experiment will have to be discarded.
To infer the weak value with a mean square precision $\Delta_w < 1$
will require on the order of $(z^2+1)/\Delta_w^2$ experimental runs.

Since the postselected estimate of $\so_z^{\rm ancilla}$ given by
(\ref{postselection_rate}) must obviously be in the range $[-1,1]$,
we see that it is necessary that $|\epsilon z| < 1$. A choice of
parameters such that $|\epsilon z| > 1$
implies that the expansion in $\epsilon$ given by
(\ref{weak_probs}) can no longer be a valid approximation. In fact, the
AAV equation (\ref{postselected_estimate}) is always meant to hold in 
the asymptotic weak measurement limit $\epsilon\rightarrow0$. 

\section{Three qubit consistency and simultaneity test of weak values}

In order to test the consistency of weak values, we need a three-qubit
quantum computer.  We use the third qubit as a second ancilla, prepared
in the initial state $\ket{\psi_{A_2}} = \cos(\vartheta_2/2)\ket{0} +
\sin(\vartheta_2/2)\ket{1}$, and perform another 
indirect weak measurement of $\so_z$ on the first qubit. The question is,
are both weak measurements {\it consistent}---that is, do both 
measurements give the same weak value ${}_f\langle\so_z\rangle_i$?

To answer this question, we perform an additional CNOT operation
between the system and the second ancilla,
to obtain the three qubit state
\begin{eqnarray}
\ket{\Psi_0} &=& \ket{\phi_i}\ket{\psi_{A_1}}\ket{\psi_{A_2}} \\
&& \longrightarrow  \ket{\Psi_{zz}} = U_{SA_2}U_{SA_1}\ket{\Psi_0}  \nonumber
\label{three_qubit_state}
\end{eqnarray}
(similar to expression (\ref{post_interaction})), with
\begin{eqnarray}
\ket{\Psi_{zz}} = \alpha \ket0 &\otimes&
    \left(\cos\frac{\vartheta_1}{2}\ket{0}+\sin\frac{\vartheta_1}{2}\ket{1}\right) \nonumber\\
&\otimes& \left(\cos\frac{\vartheta_2}{2}\ket{0}+\sin\frac{\vartheta_2}{2}\ket{1}\right) \nonumber\\
+ \beta \ket1 &\otimes& \left(\sin\frac{\vartheta_1}{2}\ket{0}+\cos\frac{\vartheta_1}{2}\ket{1}\right) \nonumber\\
&\otimes& \left(\sin\frac{\vartheta_2}{2}\ket{0}+\cos\frac{\vartheta_2}{2}\ket{1}\right) .
\label{post_interaction_2}
\end{eqnarray}

Now we count the number of events corresponding to the $\ket0$
and $\ket1$ states of both ancillas, and perform the postselection
with respect to the final system state $\ket{\phi_f}$. As we will now show,
in the weak measurement limit ($\vartheta_i=(\pi/2)-\epsilon_i, i=1,2$)
the estimates for the postselected $\so_z$ are entirely consistent.
For the probabilities, we find to leading order in $\epsilon_1,\epsilon_2$
\begin{eqnarray}
p_{f00} &\approx& \frac{|\bra{\phi_f}\phi_i\rangle|^2 
 + (\epsilon_1+\epsilon_2) \re[\bra{\phi_f}\so_z\ket{\phi_i}
                            \bra{\phi_i}\phi_f\rangle] }{4} ,\nonumber\\
p_{f01} &\approx& \frac{ |\bra{\phi_f}\phi_i\rangle|^2 
 + (\epsilon_1-\epsilon_2) \re[\bra{\phi_f}\so_z\ket{\phi_i}
                            \bra{\phi_i}\phi_f\rangle] }{4},\nonumber\\
p_{f10} &\approx& \frac{|\bra{\phi_f}\phi_i\rangle|^2 
 - (\epsilon_1-\epsilon_2) \re[\bra{\phi_f}\so_z\ket{\phi_i}
                            \bra{\phi_i}\phi_f\rangle] }{4},\nonumber\\
p_{f11} &\approx& \frac{|\bra{\phi_f}\phi_i\rangle|^2 
 - (\epsilon_1+\epsilon_2) \re[\bra{\phi_f}\so_z\ket{\phi_i}
                            \bra{\phi_i}\phi_f\rangle] }{4}. \nonumber
\label{weak_probs_2}
\end{eqnarray}
The postselected estimates of each ancilla can be determined by
averaging over the results for the other.  So we get
$p_{f0*}= p_{f00}+p_{f01}$ and $p_{f1*}= p_{f10}+p_{f11}$,
and similar expressions for $p_{f*0}$ and $p_{f*1}$.
Putting these expressions together, we get postselected expectations
that are entirely consistent with our first result
(\ref{postselected_probabilities}):
\begin{eqnarray}
\frac{p_{f0*} - p_{f1*}}{p_{f0*} + p_{f1*}} 
 & \approx &
 \epsilon_1\,\cdot\, {}_f\langle\so_z\rangle_i \nonumber\\
\frac{p_{f*0} - p_{f*1}}{p_{f*0} + p_{f*1}} 
 & \approx &
 \epsilon_2\,\cdot\, {}_f\langle\so_z\rangle_i.
\label{postselected_probabilities_2}
\end{eqnarray}
We conclude that an identical, second weak measurement of the same observable
gives the same weak value, and hence that the weak value measurements
are consistent.  Given that the correct measurement outcome is observed for
the system, all of the ancillas which interacted weakly with the system will
yield the same weak value.

A possibly even more intriguing property of weak measurement is that it
is possible to simultaneously measure consistent weak values of
{\it non-commuting} observables. To demonstrate this on a three-qubit quantum
computer, we choose to weakly measure $\so_z$ with the help of the first ancilla,
as before. The second ancilla, however, will now be used to weakly measure
$\so_x$, as briefly described at the end of the Sec.~II.

As before, we start with a three-qubit product state; but now the third
qubit is prepared in state $\ket{\psi^x_{A_2}}$,
as given in (\ref{initial_ancilla_x}). The usual CNOT$=$CNOT$^z$ operation is
performed between the system and first ancilla qubit, followed by a
CNOT$^x$ operation between the system and second ancilla qubit.
Let us now denote the unitary operation between the system and first
ancilla by $U^z_{SA_1}$ and the unitary operation between the system
and the second ancilla by $U^x_{SA_2}$.
Similar to the previous double operation (\ref{three_qubit_state}), we
obtain the three qubit state
\begin{eqnarray}
\ket{\Psi_0} &=& \ket{\phi_i}\ket{\psi^z_{A_1}}\ket{\psi^x_{A_2}}  \\
 &&  \longrightarrow \ket{\Psi_{xz}} 
      = U^x_{SA_2}U^z_{SA_1}\ket{\Psi_0} , \nonumber
\label{three_qubit_state_2}
\end{eqnarray}
with a somewhat lengthy expression for $\ket{\Psi_{xz}}$ (which we omit
for the sake of brevity).  One can think of this as weakly measuring $\so_z$
with the first ancilla and $\so_x$ with the second ancilla.  Because these are
weak measurements, this does not violate the usual restriction against
simultaneously measuring noncommuting observables, because each
individual weak measurement yields only a small amount of information.
To find the expectations, we repeat this procedure many times.

Now we count the number of events corresponding to the $\ket0$
and $\ket1$ states of the first ancilla, and the $\ket+$
and $\ket-$ state of the second ancilla.  Again, we postselect
with respect to the final system state $\ket{\phi_f}$. After some
algebra, in the weak measurement limit ($\vartheta_i=(\pi/2)-\epsilon_i, i=1,2$)
we find to leading order in $\epsilon_1,\epsilon_2$ the
probabilities
\begin{eqnarray}
p_{f0+} \approx &\frac{1}{4}& \biggl( |\bra{\phi_f}\phi_i\rangle|^2 
 + \epsilon_1 \re[\bra{\phi_f}\so_z\ket{\phi_i}
                            \bra{\phi_i}\phi_f\rangle] \nonumber \\
&& + \epsilon_2 \re[\bra{\phi_f}\so_x\ket{\phi_i}
                            \bra{\phi_i}\phi_f\rangle] \biggr),\nonumber\\
p_{f0-} \approx &\frac{1}{4}& \biggl( |\bra{\phi_f}\phi_i\rangle|^2
 + \epsilon_1 \re[\bra{\phi_f}\so_z\ket{\phi_i}
                            \bra{\phi_i}\phi_f\rangle]  \nonumber \\ 
&& - \epsilon_2 \re[\bra{\phi_f}\so_x\ket{\phi_i}
                            \bra{\phi_i}\phi_f\rangle] \biggr),\nonumber\\
p_{f1+} \approx &\frac{1}{4}& \biggl( |\bra{\phi_f}\phi_i\rangle|^2
 - \epsilon_1 \re[\bra{\phi_f}\so_z\ket{\phi_i}
                            \bra{\phi_i}\phi_f\rangle]  \nonumber \\
&& + \epsilon_2 \re[\bra{\phi_f}\so_x\ket{\phi_i}
                            \bra{\phi_i}\phi_f\rangle] \biggr),\nonumber\\
p_{f1-} \approx &\frac{1}{4}& \biggl( |\bra{\phi_f}\phi_i\rangle|^2
 - \epsilon_1 \re[\bra{\phi_f}\so_z\ket{\phi_i}
                            \bra{\phi_i}\phi_f\rangle]  \nonumber \\
&& - \epsilon_2 \re[\bra{\phi_f}\so_x\ket{\phi_i}
                            \bra{\phi_i}\phi_f\rangle] \biggr) .
\label{weak_probs_3}
\end{eqnarray}
We expect that the counts of the first ancilla will give a
postselected estimate of $\so_z$, while the counts of the
second ancilla will give a postselected estimate of $\so_x$.
With the notation
$p_{f0*}= p_{f0+}+p_{f0-}$, 
$p_{f1*}= p_{f10}+p_{f11}$, 
$p_{f*+}= p_{f0+}+p_{f1+}$, and 
$p_{f*-}= p_{f0-}+p_{f1-}$, we find
the {\it simultaneously} valid expressions 
\begin{eqnarray}
\frac{p_{f0*} - p_{f1*}}{p_{f0*} + p_{f1*}} 
 & \approx &
 \epsilon_1\,\cdot\, {}_f\langle\so_z\rangle_i \nonumber\\
\frac{p_{f*+} - p_{f*-}}{p_{f*+} + p_{f*-}} 
 & \approx &
 \epsilon_2\,\cdot\, {}_f\langle\so_x\rangle_i.
\label{postselected_probabilities_3}
\end{eqnarray}
Again, these results are entirely consistent with our first result
(\ref{postselected_probabilities}).
We conclude that the simultaneous weak measurement of non-commuting
observables gives consistent weak values for both observables.

Unsurprisingly, the order of the weak interactions is entirely
irrelevant. If we choose instead to first interact with the third and
then with the second qubit, we will obtain a (slightly) different state
$\ket{\Psi_{zx}}$; however, despite this difference, the probabilities
determined with $\ket{\Psi_{zx}}$ still coincide with the expressions 
(\ref{weak_probs_3}) above, and yield the very same postselected
estimates (\ref{postselected_probabilities_3}) as with  $\ket{\Psi_{xz}}$.

\section{Two qubit dynamical test of the weak field}
The role of the weak value in the dynamic effect on the {\it probe} was already discussed 
in Ref.~\cite{AB}. We will now show that, using just two qubits, we can
get perfect quantitative evidence of the weak value as an objective dynamic quantity
of the usual sense.

Suppose we prepare our first qubit in state $\ket{\phi_i}$ and postselect it in state 
$\ket{\phi_f}$. Between pre- and postselection we let it interact with a probe 
prepared in a certain state $\ket{\psi}$. Assume that their interaction 
Hamiltonian is $\so_z\otimes\mo$ where we can say that $\so_z$ stands for the 
``magnetic field'' of the qubit and $\mo$ stands for the ``magnetic dipole'' 
of the probe. The interaction is switched on for a short period $\delta t$ between 
pre- and postselection, and we assume that the effective
coupling remains weak.  (Its weakness will be specified later.)
We can calculate the unnormalized final state of the probe on the postselected 
statistics:
\begin{equation}
\ket{\psi}\longrightarrow \ket{\psi}-i\delta t\frac{\bra{\phi_f}\so_z\ket{\phi_i}}
                          {\braket{\phi_f}{\phi_i}}\mo\ket{\psi},
\label{qubit_probe_inter}
\end{equation} 
which means the probe feels an effective ``magnetic field''
\begin{equation}
\frac{\bra{\phi_f}\so_z\ket{\phi_i}}
                         {\braket{\phi_f}{\phi_i}}.
\label{mean_field}
\end{equation}
This quantity is complex, in general. Its real part is the weak value ${}_f\langle\so_z\rangle_i$ 
of the qubit ``magnetic field'' $\so_z$, which we could infer by doing the corresponding 
weak measurements as in Secs.~III and IV. Now we see an alternative approach:  
instead of inferring the weak value from a weak measurement, we can detect it dynamically, 
since the postselected qubit has effectively created a ``magnetic field'' 
which is equal to the weak value ${}_f\langle\so_z\rangle_i$.

The weak value anomaly is also persistent dynamically (in the same sense that it 
is robust under multiple weak measurements).  The mechanism is a natural extension 
of the usual mean-field mechanism to the case of postselection.  The surprising consequence 
in the use of postselected states is that the mean field of the qubit can be many times 
larger than the common (i.e., not postselected) mean field.
Note, however, that the interpretation requires either a weak field or a short interaction time; that is, the condition $\delta t \vert{}_f\langle\so_z\rangle_i\vert\ll1$.  To produce such a ``multiplied field'' over a longer time would require repeated postselected measurements, so that the probability of success quickly goes to zero.

We have restricted ourselves to the interpretation of the real part of the effective 
field (\ref{mean_field}).  The imaginary part is a separate issue, perhaps corresponding 
to a non-dynamical irreversible effect, superimposed on the purely dynamical effect of the real part.  
The interpretation deserves further investigation, cf. e.g. \cite{Joz07}.

Realizing such a dynamical test is straightforward. For the pre- and postselected qubit, 
we choose the example (\ref{i_f_example}), which we have already shown to produce arbitrarily 
large weak values $z$. Let the probe be a second qubit of dipole moment $\mo=\so_x$ and initial
state $\ket{\psi}=\ket{0}$. We must perform the following weak interaction:
\begin{equation}
\ket{\phi}\otimes\ket{0}\longrightarrow (1-i\delta t \so_z\otimes\so_x)\ket{\phi}\otimes\ket{0} .
\end{equation}
This is not a standard quantum-logical operation, but it should certainly be realizable by the
hardware of a quantum computer. The effect (\ref{qubit_probe_inter}) of this interaction on the probe is this:
\begin{equation}
\ket{\psi}\longrightarrow (1-i\delta t z \so_x)\ket{\psi},
\end{equation}
as if the first qubit creates a ``mean field'' $z$, and rotates the probe qubit from state $\ket{0}$ 
into $\ket{0}-i\delta t z \so_x\ket{0}$. We could choose, e.g., $z=100$ and $\delta t=1/1000$, measure 
the state of the probe in the computational basis, and detect the rate of $\ket{1}$ outcomes.  It must be $\sim(\delta tz)^2=1/100$, corresponding to a two-order-of-magnitude enhancement of the mean field.
However, we recall that $(z^2+1)/\Delta_w^2 \sim 10,000$ experimental runs 
would be needed to confirm the anomalous value z=100 by weak measurements
(see Sec.~III).  Since the dynamical effect of the post-selected qubit remains
perturbative, we would need even higher statistics to confirm the enhanced
value $z=100$ of the post-selected ``magnetic field''---approximately $10^6$ runs in the case described above.

Of course, the coupling to the probe qubit could also be added to the 
weak measurement device of Secs.~III and IV, and the dynamical effect of the post-selected qubit 
will turn out to be consistent with the outcome of the
weak measurement. Moreover, we could implement further probe qubits with 
different couplings $\mo$, that would all ``feel'' the same mean field $_i\langle\so_z\rangle_f$.

\section{Summary} 
We have presented a detailed analysis of a realistic scheme to probe the
concept of ``weak value'' in quantum mechanics \cite{AAV}, based on a quantum
computer of just two or three qubits. This seems to be in comparatively easy
reach of current quantum technology \cite{schmidtkaler2003,seidelin06,zhu2006}.
We have also discussed the appearance of the weak value anomaly, which is measurable with
just two qubits. It is possible, as we show, to test both the consistency of weak values and
the simultaneity of weak values of non-commuting observables
using three qubits. Finally, the dynamic implications of a weak value 
``mean field'' were analyzed.

We strongly believe that the realization
of a weak measurement in such a few-qubit quantum system will help to
clarify the true meaning and relevance of the concepts surrounding
the ``weak value'' in quantum mechanics. We look forward to 
seeing our proposal implemented in existing quantum computers.

\begin{acknowledgments} 
The authors would like to thank Hartmut H\"affner and Thomas Seligman 
for useful conversations. LD and WTS are grateful to the
Centro Internacional de Ciencias in Cuernavaca, Mexico, where part of
this work took shape.
TAB was supported in part by NSF Grant No.~CCF-0448658,
and LD was supported by the Hungarian Scientific Research Fund under Grant No.~49384.
\end{acknowledgments}

\end{document}